\documentclass[a4paper,amsmath,amssymb,amsfonts]{jpconf}
\usepackage{graphicx}

\begin{document}
\title{Late-time cosmic acceleration: Dark gravity}

\author{Francisco S.~N.~Lobo}

\address{Centro de Astronomia
e Astrof\'{\i}sica da Universidade de Lisboa, Campo Grande, Ed. C8
1749-016 Lisboa, Portugal}

\ead{flobo@cii.fc.ul.pt}

\begin{abstract}
A central theme in cosmology is the perplexing fact that the Universe is
undergoing an accelerating expansion. The latter, one of the most important and challenging current problems in cosmology, represents a new imbalance in the governing gravitational field equations. Several candidates, responsible for this expansion, have been proposed in the literature, in particular, dark energy models and modified gravity, amongst others. In this paper, we explore the possibility that the late-time cosmic acceleration is due to infra-red modifications of Einstein's theory of General Relativity, and review some of the modified theories of gravity that address this intriguing and exciting problem facing modern physics.
\end{abstract}

\section{Introduction}
Modern astrophysical and cosmological models are faced with two severe theoretical difficulties, that can be summarized as the dark energy and the dark matter problems. Relative to the former, high-precision observational data have confirmed with startling evidence that the Universe is undergoing a phase of accelerated expansion \cite{expansion}. Several candidates, responsible for this expansion, have been proposed in the literature, in particular, dark energy models and modified gravity, amongst others. The cause of this acceleration still remains an open and tantalizing question.
Nevertheless, a very promising way to explain these major problems is to assume that at large scales Einstein's theory of General Relativity breaks down, and a more general action describes the gravitational field. The Einstein field equation of General Relativity was first derived from an action principle by Hilbert, by adopting a linear function of the scalar curvature, $R$, in the gravitational Lagrangian density.

However, there are no a priori reasons to restrict the gravitational Lagrangian to this form, and indeed several generalizations of the Einstein-Hilbert Lagrangian have been proposed, such as involving an
arbitrary function of the scalar invariant, $f(R)$ \cite{review}.
Recently, a renaissance of $f(R)$ modified theories of gravity has
been verified in an attempt to explain the late-time accelerated
expansion of the Universe. In particular, it was shown that the cosmic acceleration can indeed be explained in the context of $f(R)$ gravity
\cite{Carroll:2003wy}.
In addition to this, in considering alternative higher-order gravity theories, one is liable to be motivated in pursuing models consistent and inspired by several candidates of a fundamental theory of quantum gravity.
Indeed, motivations from string/M-theory predict that
scalar field couplings with the Gauss-Bonnet invariant $G$
are important in the appearance of non-singular early time
cosmologies. It is also possible to apply these motivations to the
late-time Universe in an effective Gauss-Bonnet dark energy model
\cite{Nojiri:2005vv} (many aspects of Gauss-Bonnet gravity have been analyzed in the literature \cite{modGB1}).

Still in the context of modified gravity, an interesting
possibility is the existence of extra dimensions. It is widely
believed that string theory is moving towards a viable quantum
gravity theory, and one of the key predictions of string theory is
precisely the existence of extra spatial dimensions. In the
brane-world scenario, motivated by recent developments in string
theory, the observed 3-dimensional universe is embedded in a
higher-dimensional spacetime~\cite{Maartens1}.
The Dvali-Gabadadze-Porrati (DGP) models \cite{DGP} achieve a covariant infra-red modification of General Relativity, where it is possible for extra-dimensional gravity to dominate at low energies, via a brane induced gravity effect. The generalization of the DGP
models to cosmology lead to late-accelerating cosmologies
\cite{Deffayet}, even in the absence of a dark energy field
\cite{Maartens3}. This exciting feature of ``self acceleration''
may help towards a new resolution to the dark energy problem,
although this model deserves further investigation as a viable
cosmological model \cite{Lue}. While the DGP braneworld offers an
alternative explanation to the standard cosmological model, for
the expansion history of the universe, it offers a paradigm for
nature fundamentally distinct from dark energy models of cosmic
acceleration, even those that perfectly mimic the same expansion
history. It is also fundamental to understand how one may
differentiate this modified theory of gravity from dark energy
models.

In this work, we review several modified theories of gravity, namely, $f(R)$ gravity and the DGP braneworld model, briefly exploring some of their interesting properties and characteristics, and in particular, focussing mainly on the late-time cosmic acceleration.

\section{Modified theories of gravity: Late-time cosmic acceleration}
\label{sec:II}

\subsection{$f(R)$ modified theories of gravity}\label{sec:IIa}

A promising avenue that has been extensively investigated recently
are the $f(R)$ modified theories of gravity, where the standard
Einstein-Hilbert action is replaced by an arbitrary function of
the Ricci scalar $R$.

The action for the $f(R)$ modified theories of gravity is given by
\begin{equation}
S=\frac{1}{2\kappa}\int d^4x\sqrt{-g}\;f(R)+S_M(g^{\mu\nu},\psi)
\,,
\end{equation}
where $\kappa =8\pi G$. $S_M(g^{\mu\nu},\psi)$ is the matter
action, in which matter is minimally coupled to the
metric $g_{\mu\nu}$ and $\psi$ collectively denotes the matter
fields. Varying the action with respect to $g^{\mu\nu}$, provides the following field equation
\begin{equation}
FR_{\mu\nu}-\frac{1}{2}f\,g_{\mu\nu}-\nabla_\mu \nabla_\nu
F+g_{\mu\nu}\nabla_\alpha \nabla^\alpha F=\kappa\,T^{(m)}_{\mu\nu} \,,
    \label{field:eq}
\end{equation}
where $F\equiv df/dR$.

It is interesting to note that $f(R)$ gravity may lead to an
effective dark energy, without the need to introduce a negative
pressure ideal fluid. Taking into account the flat FRW metric and a perfect fluid description for matter, we verify that the gravitational field equation, Eq. (\ref{field:eq}), provides the generalised Friedmann equations in the following form \cite{Capozziello:2003tk}:
\begin{eqnarray}
\left(\frac{\dot{a}}{a}\right)^2=\frac{\kappa}{3}\rho_{\rm tot} \,,  \qquad
\left(\frac{\ddot{a}}{a}\right)=-\frac{\kappa}{6}(\rho_{\rm
tot}+3p_{\rm tot}) \,,
    \label{rho+3p}
\end{eqnarray}
where $\rho_{\rm tot}=\rho+\rho_{(c)}$ and $p_{\rm
tot}=p+p_{(c)}$, and the curvature stress-energy components,
$\rho_{(c)}$ and $p_{(c)}$, are defined as
\begin{eqnarray}
\rho_{(c)}&=&\frac{1}{\kappa F(R)}\left\{\frac{1}{2}
\left[f(R)-RF(R)\right]-3\left(\frac{\dot{a}}{a}\right)\dot{R}
F'(R)\right\} \,,  \\
p_{(c)}&=&\frac{1}{\kappa
F(R)}\left\{2\left(\frac{\dot{a}}{a}\right)\dot{R}
F'(R)+\ddot{R}F'(R)+\dot{R}^2F''(R)-\frac{1}{2}\left[f(R)-RF(R)\right]\right\}
 \,,
\end{eqnarray}
respectively. The late-time cosmic acceleration is achieved if the
condition $\rho_{\rm tot}+3p_{\rm tot}<0$ is obeyed, which follows
from Eq. (\ref{rho+3p}).

For simplicity, consider the absence of matter, $\rho=p=0$. Now,
taking into account the equation of state $\omega_{\rm
eff}=p_{(c)}/\rho_{(c)}$, with $f(R)\propto R^n$ and a generic
power law $a(t)=a_0(t/t_0)^\alpha$ \cite{Capozziello:2003tk}, the
parameters $\omega_{\rm eff}$ and $\alpha$ are given by
\begin{equation}
\omega_{\rm eff}=-\frac{6n^2-7n-1}{6n^2-9n+3} \,, \qquad
\alpha=\frac{-2n^2+3n-1}{n-2} \,,
\end{equation}
respectively, for $n\neq 1$. Note that a suitable choice of $n$
can lead to the desired value of $\omega_{\rm eff}<-1/3$,
achieving the late-time cosmic acceleration.

Other forms of $f(R)$ have also been considered in the literature,
for instance those involving logarithmic terms \cite{modGB1}. These models also yield acceptable values for the effective equation of state parameter, resulting in the late-time cosmic acceleration.

\subsection{DGP brane gravity and self-acceleration}\label{sec:IIc}

One of the key predictions of string theory is the existence of
extra spatial dimensions. In the brane-world scenario, motivated
by recent developments in string theory, the observed
3-dimensional universe is embedded in a higher-dimensional
spacetime~\cite{Maartens1}. One of the simplest covariant models
is the Dvali-Gabadadze-Porrati (DGP) braneworld model, in which
gravity leaks off the $4D$ Minkowski  brane into the $5D$ bulk at
large scales. The generalization of the DGP models to cosmology
lead to late-accelerating cosmologies \cite{Deffayet}, even in the
absence of a dark energy field \cite{Maartens3}.

The $5D$ action describing the DGP model is given by
\begin{equation}
S=\frac{1}{2\kappa_5}\int
d^5x\sqrt{-g}\;^{(5)}R+\frac{1}{2\kappa_4}\int
d^4x\sqrt{-\gamma}\;^{(4)}R-\int d^4x\sqrt{-\gamma}\;{\cal L}_m
\,,
\end{equation}
where $\kappa_5=8\pi G_5$ and $\kappa_4=8\pi G_4$. The first term
in the action is the Einstein-Hilbert action in five dimensions
for a five-dimensional bulk metric, $g_{AB}$, with a
five-dimensional Ricci scalar $^{(5)}R$; the second term is the
induced Einstein-Hilbert term on the brane, with a
four-dimensional induced metric $\gamma$ on the brane; and ${\cal
L}_m$ represents the matter Lagrangian density confined to the
brane.

The transition from $4D$ to $5D$ behavior is governed by a
cross-over scale, $r_c$, given by $r_c=\frac{\kappa_5}{2\kappa_4}$.
Gravity manifests itself as a $4$-dimensional theory for
characteristic scales much smaller than $r_c$; for large distances
compared to $r_c$, one verifies a leakage of gravity into the
bulk, consequently making the higher dimensional effects
important. Thus, the leakage of gravity at late times initiates
acceleration, due to the weakening of gravity on the brane.

Taking into account the FRW metric, and considering a flat
geometry, the modified Friedmann equation is given by
\begin{equation}
H^2-\frac{\epsilon}{r_c}H=\frac{8\pi G}{3}\rho\,,
   \label{DGPfriedeq}
\end{equation}
where $\epsilon=\pm 1$, and the energy density satisfies the
standard conservation equation, i.e., $\dot{\rho}+3H(\rho+p)=0$.
For scales $H^{-1}\ll r_c$, the second term is negligible, and Eq.
(\ref{DGPfriedeq}) reduces to the general relativistic Friedmann
equation, i.e., $H^2=8\pi G\rho/3$. The second term becomes
significant for scales comparable to the cross-over scale,
$H^{-1}\geq r_c$. Self-acceleration occurs for the branch
$\epsilon =+1$, and the modified Friedmann equation shows that at
late times in a CDM universe characterized by a scale factor $\rho
\propto a^{-3}$, the universe approaches a de Sitter solution
$H\rightarrow H_{\infty}=\frac{1}{r_c}$.
Thus, one may achieve late-time acceleration if the $H_0$ is of
the order of $H_{\infty}$. Note that the late-time acceleration in
the DGP model is not due to the presence of a negative pressure,
but simply due to the weakening of gravity on the brane as a
consequence of gravity leakage at late times.

Although the weak-field gravitational DGP behaves as $4D$ on
scales smaller than $r_c$, linearized DGP gravity is not described
by General Relativity \cite{Koyama:2007za}. It is
also interesting to note that despite the fact that the expansion
history of the DGP model and General Relativity are the same, the
structure formation in both are essentially different
\cite{Koyama:2005kd}. Combining these features provides the
possibility of distinguishing the DGP model from dark energy
models in General Relativity. Another interesting aspect of this model is that the self-accelerating branch in the DGP model contains a ghost at the
linearized level \cite{Koyama:2007za,Koyama:2005tx}. The presence
of the ghost implies a negative sign for the kinetic term,
resulting in negative energy densities, consequently leading to
the instability of the spacetime. We refer the
reader to Ref. \cite{Koyama:2007za}, and references
therein, for more details on the DGP model.

\section{Conclusion}

The standard model of cosmology is remarkably successful in
accounting for the observed features of the Universe. However,
there remain a number of fundamental open questions at the
foundation of the standard model. In particular, we lack a
fundamental understanding of the acceleration of the late
universe. Recent observations of supernovae, together with the
WMAP and SDSS data, lead to the remarkable conclusion that our
universe is not just expanding, but has begun to
accelerate~\cite{expansion}. What is the so-called `dark energy'
that is driving the acceleration of the universe? Is it a vacuum
energy or a dynamical field (``quintessence'')? Or is the
acceleration due to infra-red modifications of Einstein's theory
of General Relativity? How is structure formation affected in
these alternative scenarios? What will the outcome be of this
acceleration for the future fate of the universe?
Deciding between these possible sources of the cosmic acceleration
will be one of the major objectives in cosmology in the next
decade with several surveys and experiments to address the nature
of dark energy. All of these aspects present an
extremely fascinating aspect for future theoretical research.

\end{document}